\documentclass{article}
\usepackage{spconf,amsmath,amssymb,graphicx,hyperref}
\usepackage{booktabs}
\usepackage{multirow}
\usepackage{makecell}
\usepackage{array}
\usepackage{url,cite}
\usepackage[subtle,lists=normal]{savetrees}

\title{USING THE ENROLLMENT GALLERY AS EVIDENCE:\\
AFFINE-INVARIANT SCORE CALIBRATION FOR TARGET SPEAKER TAGGING}

\name{Hee-Soo Heo, Minjae Lee, Youngki Kwon, Bong-Jin Lee}
\address{NAVER Cloud Corporation, South Korea}

\begin{document}
\ninept
\maketitle

\begin{abstract}
Target speaker tagging (TST) assigns enrolled speaker identities to diarized segments of multi-speaker recordings.
The enrollment utterances of the other registered speakers are an attractive source of evidence for calibrating each verification decision: they share the deployment domain of the test material and require no external cohort set.
We show, however, that they cannot serve as the cohort of conventional score normalization.
Enrollments tend to cluster by recording session, so the per-speaker cohort statistics reflect enrollment proximity to the rest of the gallery rather than impostor behavior, and normalization then rejects entire speakers.
We propose gallery affinity verification, a cohort-free calibration that exploits the gallery through two affine-invariant terms, score dispersion and enrollment-profile deviance, and is therefore immune to such per-speaker shifts.
On a synthetic benchmark and an in-house meeting corpus, it improves tagging accuracy with and without conventional score normalization and adds further gains when combined with it.
\end{abstract}


\section{Introduction}
\label{sec:intro}

Target speaker tagging (TST)~\cite{tst2026} is a composite speaker recognition task in which a long multi-speaker recording is segmented into single-speaker regions~\cite{Campbell1997, bai2021speaker, park2022review, fujita2019end} and each segment is either assigned the identity of a pre-enrolled speaker or rejected as non-target~\cite{snyder2018x, wang2021dku, park2019auto, desplanques2020ecapa}.
TST directly supports applications such as meeting transcription and multi-session analytics, where knowing who said what is essential.
Prior work has established the task definition and benchmark~\cite{tst2026}, robust session-level score calibration~\cite{gmm2026}, and a decoupled identifier--verifier architecture that mitigates false alarms at scale~\cite{asimi2026}.

A defining property of TST service scenarios is that the system accumulates enrollment utterances for every registered speaker.
These recordings are uniquely valuable: they are captured by the same acquisition pipeline and often under the same acoustic conditions as the material to be recognized.
In a typical TST service, enrollment speech is selected from the recordings that the system processes, and the remaining utterances of those sessions are exactly the material to be tagged; enrollment and test speech are thus drawn from one shared pool rather than from two separate collections.
Exploiting them beyond their primary role of forming each speaker's reference embedding is therefore a natural direction for improving the verification performance.

Cohort-based score normalization~\cite{auckenthaler2000score, cumani2011comparison} is the standard way of incorporating auxiliary utterances into a verification decision.
A natural choice is therefore to use the other enrolled speakers' utterances as the cohort.
This cohort, however, differs from the large and diverse impostor sets that score normalization assumes.
The gallery is small, and its members are not sampled independently from an impostor population: subsets of enrolled speakers share recording channels, rooms, and conversational domains.
As we show, the per-speaker statistics computed on such a cohort measure how close a speaker's enrollments are to the rest of the gallery, not how impostors score against that speaker.
Normalizing with these statistics introduces speaker-dependent score offsets.
Scores then no longer remain comparable across speakers, a property that an identification system with a shared decision threshold requires.

This paper makes three contributions.
First, we present a systematic analysis of standard score normalization in TST under two cohort choices, quantitatively isolating why the enrollment gallery cannot serve as a normalization cohort (Sec.~\ref{ssec:norm_limits}).
Second, we propose gallery affinity verification, a cohort-free calibration that uses the gallery through two affine-invariant terms---a score dispersion term and an enrollment-profile deviance term---designed to be invariant to the per-speaker offset and scale distortions that break cohort-based normalization (Sec.~\ref{sec:proposed}).
Third, we show on two datasets that the proposed method improves the tagging accuracy both with and without conventional score normalization (Sec.~\ref{sec:experiments}).

\section{Background and Problem Analysis}
\label{sec:background}

\subsection{Target speaker tagging system}
\label{ssec:tst_system}

We follow the TST framework of~\cite{tst2026}.
During enrollment, speaker embeddings are extracted from each enrolled speaker's utterances, using 12--60 seconds of speech per speaker, and stored to form a gallery $\mathcal{G}=\{(\mathbf{e}_k,\mathrm{id}_k)\}_{k=1}^{N}$, where $\mathbf{e}_k$ denotes the window-level embeddings of the $k$-th enrolled speaker, $\mathrm{id}_k$ the corresponding identity, and $N$ the number of enrolled speakers.
At test time, a clustering-based diarization front-end (Sec.~\ref{ssec:setup}) partitions the recording into single-speaker segments carrying session-local cluster labels.

Identification and verification are decoupled following~\cite{asimi2026}: the identifier and the verifier each own a separate embedding extractor.
An extractor maps speech to speaker embeddings, and a segment is scored against an enrolled speaker by the cosine similarity between their embeddings, computed over all window-level embeddings of the two sides.
With the identifier's embeddings, each segment $t$ is scored against every enrolled speaker $k$, which yields the identifier scores $s_{t,k}$, and the top candidate $c=\arg\max_k s_{t,k}$ is selected; the verifier scores, computed from the verifier's embeddings in the same way, are then used for the verification decision on $c$.
The two extractors are distinct checkpoints of the same embedding model taken from different training stages, which provides the extractor diversity that the decoupled framework requires.
The concrete models are given in Sec.~\ref{ssec:setup}.
Many diarized segments are short, and the verifier scores of such segments are individually unreliable.
The session-level GMM calibration of~\cite{gmm2026} reduces this uncertainty by aggregating the verifier scores within each diarization cluster while remaining robust to clustering errors.
Specifically, a two-component GMM is fitted to the verifier-score distribution of each diarization label, one component capturing correctly clustered segments and the other absorbing misclustered ones, and each segment--speaker pair receives a GMM-calibrated score, the likelihood-based aggregate of its verifier scores.
A single global threshold $\theta$ on this score, shared by all enrolled speakers, determines acceptance.

Unlike the evaluation protocol of~\cite{tst2026}, which restricts each session's candidate set to the speakers known to attend that session, we place every enrolled speaker in the candidate set of every session.
This reflects deployments in which attendance information is unavailable, and makes the rejection problem strictly harder: each decision competes against the full gallery.

\subsection{Limits of the gallery as a normalization cohort}
\label{ssec:norm_limits}

Cohort-based normalization rescales the verifier score $s(e,t)$ of a trial between enrolled speaker $e$ and test segment $t$ using statistics of cohort scores.
Z-norm normalizes the enrollment side, $s'=(s-\mu_e)/\sigma_e$, where $\mu_e$ and $\sigma_e$ are the mean and standard deviation of the scores between $e$'s enrollments and the cohort; T-norm normalizes the test side analogously with $(\mu_t,\sigma_t)$; S-norm averages the Z-norm and T-norm scores; and adaptive S-norm (AS-norm) computes each side's statistics only over the top-$K$ closest cohort members~\cite{cumani2011comparison, matejka2017analysis}.

To determine whether the enrollment gallery can substitute for the external data that these methods normally require, we compare two cohort choices under otherwise identical conditions: (i)~an external cohort of 2{,}170 four-second utterances drawn from VoxBlink2~\cite{lin2024voxblink2}, matching the embedding window length (Sec.~\ref{ssec:setup}) and following the cohort choice of~\cite{tst2026, gmm2026}; and (ii)~the enrolled cohort, consisting of the enrollment embeddings of all gallery speakers except the current candidate.
All methods are applied to the verifier scores before the session-level GMM calibration, and the identifier is left unchanged.

\begin{table}[t]
\centering
\caption{Score normalization under two cohort choices
(DIR@FAR=1\%,~[\%]). ``Ext.'' is the VoxBlink2 cohort; ``Enr.'' uses the
other enrolled speakers. For AS-norm, several top-$K$ sizes were tried and the best result is reported for each cohort.}
\label{tab:norm_grid}
\setlength{\tabcolsep}{5pt}
\begin{tabular}{lcccc}
\toprule
\multirow{2}{*}{Method} & \multicolumn{2}{c}{In-house} & \multicolumn{2}{c}{TST-Bench} \\
\cmidrule(lr){2-3}\cmidrule(lr){4-5}
 & Ext. & Enr. & Ext. & Enr. \\
\midrule
No normalization & \multicolumn{2}{c}{81.64} & \multicolumn{2}{c}{87.89} \\
Z-norm  & 85.26 & 41.72 & 81.28 & 82.20 \\
T-norm  & 85.42 & 47.84 & 80.26 & 80.74 \\
S-norm  & 85.35 & 46.28 & 82.34 & 84.00 \\
AS-norm & 82.75 & 77.42 & 83.17 & 80.76 \\
\bottomrule
\end{tabular}
\end{table}

Table~\ref{tab:norm_grid} compares the two cohorts.
On the in-house corpus, the external cohort behaves as reported in the verification literature: Z-, T-, and S-norm improve DIR@FAR=1\% by roughly 4 percentage points (pp) over the unnormalized baseline.
With the enrolled cohort, however, the same methods lose roughly half of the baseline accuracy on that corpus.

To understand this failure, we examined the in-house results (the corpus is described in Sec.~\ref{ssec:setup}).
Two observations about the enrollment utterances that form the enrolled cohort stand out.
First, the similarity between two utterances, be they enrollments or test segments, is higher when they come from the same or a similar session, because the shared recording channel is reflected in the embeddings.
Second, speaker enrollments are clustered by session: multiple speakers are often enrolled from the same session, for example when a meeting recording is used to enroll its attendees.
Subsets of the gallery therefore share a recording condition.
Neither observation is specific to our corpus; both follow from how enrollment data is collected in TST deployments.

Clustered enrollment distorts the enrolled-cohort statistics.
For an enrolled speaker $e$, the enrollment-side mean $\mu_e$ is the average similarity between $e$'s enrollments and those of the other gallery speakers.
Speakers enrolled together with many others receive a large $\mu_e$, not because impostors resemble them but because their enrollments share a channel with part of the gallery.
Their test segments, recorded in other sessions, carry no such boost.
Normalization therefore subtracts from their trial scores an offset that these scores never contained, pushing all of their trials, genuine targets included, below the shared threshold; in our results, the speakers with the largest $\mu_e$ indeed lose most of their target segments.
The speakers enrolled far from the rest of the gallery receive the opposite treatment: their normalized scores become relatively the highest, and false alarms concentrate on them.
Mass rejection and false-alarm concentration thus occur together because they affect different speakers.

On an external cohort, the per-speaker statistics instead track each speaker's similarity to a broad, independent impostor population, which is the quantity normalization is designed to remove, and they vary little across speakers.
The external cohort therefore calibrates scores within each trial while preserving the ordering of speakers, whereas the enrolled cohort reorders the speakers on an erroneous basis.

TST-Bench provides a control condition for this account.
Its sessions are synthesized from MLS~\cite{pratap2020mls} audiobook recordings under nearly uniform acoustic conditions, so no subset of enrolled speakers shares a distinctive channel and the enrolled-cohort statistics are almost as uniform as external ones.
Accordingly, the enrolled cohort shows no failure there and performs on par with the external cohort (Table~\ref{tab:norm_grid}), confirming that the failure follows the clustering of enrollment conditions rather than the gallery size.
That neither cohort improves over the TST-Bench baseline has a complementary explanation.
The uniform synthetic conditions leave little cross-speaker or cross-session miscalibration to correct, and the VoxBlink2 cohort is acoustically distant from the MLS-derived audio; normalization therefore adds estimation noise without a compensating benefit.

These findings motivate a different use of the gallery: the informative quantity is the pattern of one segment's scores across the enrolled speakers, not the per-speaker score statistics.

\section{Proposed Gallery Affinity Verification}
\label{sec:proposed}

We keep the identifier's candidate decision unchanged and calibrate only the verification score of the selected candidate $c$; that is, the proposed terms are applied at the verifier, the stage at which score normalization is conventionally applied (Sec.~\ref{ssec:norm_limits}).
Let $\mathbf{s}=[s_1,\dots,s_N]$ denote the segment's GMM-calibrated scores (Sec.~\ref{ssec:tst_system}) over the $N$ enrolled speakers.
For any vector $\mathbf{x}=[x_1,\dots,x_N]\in\mathbb{R}^{N}$, we write $\mathbf{x}_{-c}\in\mathbb{R}^{N-1}$ for the vector that collects the entries $x_k$, $\forall k\neq c$, in their original order; in particular, $\mathbf{s}_{-c}$ is the segment's competitor profile, containing $m=N{-}1$ entries.
The final decision score is
\begin{equation}
D_c \;=\; s_c \;-\; \rho(\mathbf{s}_{-c}) \;-\;
\frac{\max\!\left(r_c^2-f_c^2,\,0\right)}{2 f_c},
\label{eq:decision}
\end{equation}
whose two subtracted terms are defined below: $\rho(\mathbf{s}_{-c})$ measures the dispersion of the competitor scores (Sec.~\ref{ssec:dispersion}), and the last term penalizes the deviation of the segment's competitor profile from the candidate's enrollment profile, measured by the residual $r_c$ relative to the residual level $f_c$ that is normal for candidate $c$ (Sec.~\ref{ssec:deviance}).

\subsection{Competitor-profile dispersion}
\label{ssec:dispersion}
The first penalty measures how widely the segment's scores $\mathbf{s}$ spread over the gallery.
Because the candidate is the maximum over $N$ scores, a widely spread profile is a risk in itself: even for a non-target segment, the top of a wide score distribution is likely to exceed the threshold, producing a false alarm.
The dispersion term quantifies this spread as the distance between the upper and lower ends of the competitor profile $\mathbf{s}_{-c}$ and subtracts it from the candidate score.

Formally, let $Q_q(\mathbf{s}_{-c})$ denote the linearly interpolated empirical $q$-quantile of the competitor profile, and let the tail depth $d$ be the number of profile entries treated as the tail on each side.
With the tail fraction $\tau=d/m$, the dispersion term is the symmetric interquantile radius
\begin{equation}
\rho(\mathbf{s}_{-c}) \;=\;
\tfrac{1}{2}\left[\,Q_{1-\tau}(\mathbf{s}_{-c}) -
Q_{\tau}(\mathbf{s}_{-c})\,\right].
\label{eq:dispersion}
\end{equation}
The tail depth is the only design choice of this term.
Rather than tuning it, we tie it to the gallery size by a square-root tail rule, $d=\lfloor\sqrt{m}\rfloor$, which we adopt to balance two requirements as the gallery grows: each quantile is supported by $\sqrt{m}$ entries, so its estimate remains stable, while the tail fraction $\tau=1/\sqrt{m}$ vanishes, so the radius approaches the full spread of the profile except for its $\sqrt{m}$ most extreme entries on each side.
The term therefore has no free hyperparameter.
With this choice, the statistic responds when many competitor scores are elevated at once, while remaining robust to a few competitors who happen to resemble the segment.

The reason $\rho$ separates impostors from targets is as follows.
A genuine target segment is close to its own speaker and roughly uniformly dissimilar to the rest of the gallery, so its competitor profile is narrow and $\rho$ is small.
An impostor segment that reaches a high candidate score typically does so through channel or domain proximity, and the same proximity elevates its scores against many other enrolled speakers, widening the profile and increasing $\rho$.
The penalty thus grows exactly in the situation where a high $s_c$ is least reliable.

\subsection{Enrollment-profile deviance}
\label{ssec:deviance}

At enrollment time we compute, for each speaker $k$, an enrollment profile $\mathbf{p}_k\in\mathbb{R}^{N}$: the vector of scores between $k$'s enrollment speech and every gallery speaker, obtained from enrollment audio only.
Although enrollment audio carries no diarization errors by construction, the profiles are passed through the same GMM calibration as the test-side scores, so that both profiles compared below live in the same score domain.
At test time, the segment's competitor profile is compared against the candidate's enrollment profile restricted to competitors, $(\mathbf{p}_c)_{-c}$, through a non-negative affine fit.
Writing $\bar{\mathbf{s}}$ and $\bar{\mathbf{p}}$ for the centered versions of $\mathbf{s}_{-c}$ and $(\mathbf{p}_c)_{-c}$, the fitted slope and the degree-of-freedom-corrected residual~\cite{draper1998applied} are
\begin{equation}
\alpha_c = \max\!\left(
\frac{\langle\bar{\mathbf{s}},\bar{\mathbf{p}}\rangle}
{\lVert\bar{\mathbf{p}}\rVert^2},\,0\right),
\qquad
r_c = \sqrt{\frac{\lVert\bar{\mathbf{s}}-\alpha_c\,\bar{\mathbf{p}}
\rVert^2}{m-2}}.
\label{eq:residual}
\end{equation}
Because the fit absorbs an arbitrary offset (through centering) and an arbitrary non-negative scale (through $\alpha_c$), $r_c$ measures only whether the shape of the segment's relation to the gallery matches the candidate's known relation.
It is invariant to exactly the per-speaker affine distortions identified in Sec.~\ref{ssec:norm_limits}.
The slope is clamped at zero so that an anti-correlated competitor profile receives no credit from the fit.

The residual $r_c$ alone, however, is not comparable across candidates.
Even one speaker's own enrollment utterances deviate from that speaker's profile, and the size of this deviation differs between speakers.
A fixed penalty on $r_c$ would thus punish speakers with less stable profiles even for genuine segments.
We instead compare $r_c$ with a per-speaker reference level $f_c$ obtained at enrollment time as follows.
For each enrollment utterance of $c$, its profile is computed in the same way as for a test segment.
Each such profile is fitted to $(\mathbf{p}_c)_{-c}$ with Eq.~(\ref{eq:residual}), giving one residual per utterance.
Their median is $f_c$, the residual a genuine utterance of $c$ typically attains.
The penalty in Eq.~(\ref{eq:decision}), $\max(r_c^2-f_c^2,0)/(2f_c)$, is a smooth hinge around this reference.
It is zero while $r_c$ stays within the range observed during enrollment, close to $r_c - f_c$ for mild excesses, and quadratic for strong mismatches.
Because $f_c$ comes from the candidate's own enrollment data, this term has no hyperparameter either.

The deviance term checks whether the segment relates to the other enrolled speakers in the same way as the claimed speaker's own enrollments do.
For a genuine target this holds by construction, since the same voice generated both profiles, so $r_c\!\approx\!f_c$ and no penalty is applied, regardless of any global score shift the segment may carry.
An impostor accepted as candidate $c$, in contrast, carries the affinity pattern of its own identity and channel.
No offset and non-negative scale maps its competitor profile onto $(\mathbf{p}_c)_{-c}$, so $r_c\!\gg\!f_c$ and the decision score is reduced.

\subsection{Relation to score normalization}
\label{ssec:relation}
Both cohort normalization and the proposed terms draw on the same gallery, which raises an immediate question: why does the clustering of enrollment conditions that breaks the former (Sec.~\ref{ssec:norm_limits}) not break the latter?
In cohort-based normalization, the statistics $(\mu,\sigma)$ computed from the cohort---here, the enrollment utterances---enter the normalized score directly.
Any characteristic of the cohort, such as clustered recording conditions, therefore propagates unchanged into the corrected scores and shifts entire speakers.
The proposed method uses the same utterances only to form similarity patterns that are compared between enrollment and test through a non-negative affine fit.
The fit removes exactly the per-speaker offset and scale through which such characteristics distort the scores.
The dispersion term consults no enrollment statistics at all and relies only on the competitor scores of the current segment.
Under this comparison the clustering becomes evidence rather than a nuisance: enrollments recorded under conditions shared with part of the gallery give the candidate a distinctive profile, which a genuine segment of that speaker reproduces and an impostor does not.
The two families are therefore complementary: normalization calibrates the scale of each score against a representative impostor population, whereas the proposed terms test whether the pattern of scores is consistent with the gallery structure.
Sec.~\ref{ssec:main_results} confirms this empirically.

\section{Experiments}
\label{sec:experiments}

\subsection{Setup}
\label{ssec:setup}

\textbf{Datasets.} We evaluate on (i)~TST-Bench~\cite{tst2026}, a synthetic benchmark built from MLS~\cite{pratap2020mls} with 150 enrolled speakers, 300 sessions, and 20{,}252 evaluated segments; and (ii)~an in-house Korean meeting corpus of 84 sessions with 53 enrolled speakers (16{,}677 target and 5{,}061 non-target segments), covering team meetings and small-group conversations.
In both cases the full gallery serves as the candidate set for every session (Sec.~\ref{ssec:tst_system}).

\textbf{System.} Both the identifier and the verifier use the ReDimNet2 architecture in its B6 configuration~\cite{yakovlev2024reshape, yakovlev2026redimnet2}, trained on VoxBlink2 and VoxCeleb2~\cite{lin2024voxblink2, Nagrani17, chung2018voxceleb2, Nagrani19}.
The identifier uses the pre-trained checkpoint and the verifier its large-margin fine-tuned counterpart\footnote{Both checkpoints are available at \url{https://github.com/PalabraAI/redimnet2}.}.
Embeddings are extracted with 4-second windows and 1.5-second shifts.
Diarized segments are produced by a clustering-based diarization system built on the high-resolution embedding extractor of~\cite{heo2023high}.

\textbf{Metrics.} Following~\cite{tst2026, lin2024voxblink2, jain2011handbook}, we report the detection and identification rate (DIR) at fixed false alarm rates: a target segment counts as correct only if it is accepted with the correct identity, and a non-target segment counts as a false alarm if it is accepted at all.
Because each experimental condition alters the decision-score domain, its threshold is re-tuned to each target FAR on the per-segment decision scores, and we report DIR at the matched achievable FAR.
We additionally report the equal error rate (EER), which summarizes the full trade-off curve and complements the fixed-FAR operating points.

\subsection{Main results}
\label{ssec:main_results}

\begin{table}[t]
\centering
\caption{Main results: the proposed calibration with and without
score normalization (external cohort; AS-norm with $K{=}100$). DIR@FAR and EER in \%.}
\label{tab:main}
\setlength{\tabcolsep}{3.2pt}
\resizebox{\columnwidth}{!}{%
\begin{tabular}{llccccc}
\toprule
\multirow{2}{*}{Dataset} & \multirow{2}{*}{Method} & \multicolumn{4}{c}{DIR@FAR [\%]} & \multirow{2}{*}{EER} \\
\cmidrule(l){3-6}
 & & 0.5\% & 1\% & 2\% & 5\% & \\
\midrule
\multirow{6}{*}{TST-Bench}
 & Baseline             & 82.17 & 87.89 & 90.00 & 92.23 & 7.03 \\
 & ~~$+$ Proposed       & \textbf{87.06} & \textbf{91.10} & \textbf{92.66} & \textbf{93.82} & \textbf{5.98} \\
 & AS-norm              & 77.92 & 83.17 & 86.84 & 90.08 & 8.25 \\
 & ~~$+$ Proposed       & 83.39 & 89.39 & 91.51 & 93.39 & 6.35 \\
 & T-norm               & 75.54 & 80.26 & 84.47 & 87.75 & 9.61 \\
 & ~~$+$ Proposed       & 81.15 & 87.62 & 90.33 & 92.56 & 6.87 \\
\midrule
\multirow{6}{*}{In-house}
 & Baseline             & 79.95 & 81.64 & 84.07 & 87.65 & 11.07 \\
 & ~~$+$ Proposed       & \textbf{81.78} & 85.25 & 87.05 & 88.48 & 10.00 \\
 & AS-norm              & 77.88 & 82.75 & 84.16 & 86.21 & 11.66 \\
 & ~~$+$ Proposed       & 81.23 & 84.94 & 87.63 & 90.75 & 8.66 \\
 & T-norm               & 80.75 & 85.42 & 87.40 & 88.39 & 9.54 \\
 & ~~$+$ Proposed       & 81.32 & \textbf{87.72} & \textbf{89.08} & \textbf{91.40} & \textbf{8.08} \\
\bottomrule
\end{tabular}}
\end{table}

Table~\ref{tab:main} presents the main results.
Applied on top of the unnormalized baseline, the proposed calibration improves every operating point on both datasets, with gains of roughly $3$--$5$\,pp, achieved without any external cohort.
Applied on top of external-cohort normalization, it improves every operating point as well and yields the best EER in each block.
The gallery evidence exploited by the proposed terms is thus complementary to cohort-based calibration rather than redundant with it.
Overall, the method helps when normalization is unavailable, and it helps further when normalization is already in place.

\subsection{Ablation: contribution of each term}
\label{ssec:ablation}

\begin{table}[t]
\centering
\caption{Term-level ablation on TST-Bench (DIR@FAR~[\%]).}
\label{tab:ablation}
\setlength{\tabcolsep}{4pt}
\begin{tabular}{lcccc}
\toprule
\multirow{2}{*}{Decision score} & \multicolumn{4}{c}{DIR@FAR [\%]} \\
\cmidrule(l){2-5}
 & 0.5\% & 1\% & 2\% & 5\% \\
\midrule
$s_c$ (baseline)         & 82.17 & 87.89 & 90.00 & 92.23 \\
$+$ dispersion           & 83.67 & 88.77 & 90.35 & 92.60 \\
$+$ deviance             & \textbf{87.42} & 90.71 & 92.44 & 93.72 \\
$+$ both (proposed)      & 87.06 & \textbf{91.10} & \textbf{92.66} & \textbf{93.82} \\
\bottomrule
\end{tabular}
\end{table}

Table~\ref{tab:ablation} isolates the two terms.
This and the following ablations use TST-Bench, whose larger number of sessions and balanced speaker distribution yield more stable estimates than the in-house corpus.
Both terms contribute individually: at FAR=1\%, the dispersion term alone improves DIR by about $0.9$\,pp and the deviance term by about $2.8$\,pp, and their combination reaches $3.2$\,pp, indicating that the two terms capture partially independent evidence.
The combination is the strongest configuration at every operating point except the strictest one, where the deviance term alone is marginally better.

\subsection{Ablation: interaction with session-level calibration}
\label{ssec:gmm_ablation}

\begin{table}[t]
\centering
\caption{Effect of the proposed terms with and without the session-level
GMM calibration of~\cite{gmm2026}, on TST-Bench (DIR@FAR~[\%]).}
\label{tab:gmm_ablation}
\setlength{\tabcolsep}{4pt}
\resizebox{\columnwidth}{!}{%
\begin{tabular}{llcccc}
\toprule
\multirow{2}{*}{Session-level calibration} & \multirow{2}{*}{Method} & \multicolumn{4}{c}{DIR@FAR [\%]} \\
\cmidrule(l){3-6}
 & & 0.5\% & 1\% & 2\% & 5\% \\
\midrule
\multirow{2}{*}{None (raw scores)}
 & Baseline       & 76.76 & 82.27 & 86.22 & 89.39 \\
 & $+$ Proposed   & 81.21 & 86.08 & 87.82 & 90.71 \\
\midrule
\multirow{2}{*}{GMM~\cite{gmm2026}}
 & Baseline       & 82.17 & 87.89 & 90.00 & 92.23 \\
 & $+$ Proposed   & \textbf{87.06} & \textbf{91.10} & \textbf{92.66} & \textbf{93.82} \\
\bottomrule
\end{tabular}}
\end{table}

The session-level GMM calibration reshapes the score distribution that the proposed terms operate on.
The gains reported so far could therefore in principle be specific to GMM-calibrated scores.
To rule this out, Table~\ref{tab:gmm_ablation} repeats the comparison with the calibration removed, applying the proposed terms directly to raw cosine scores.
The proposed terms yield gains of comparable magnitude at every operating point, and the two mechanisms are additive: the strongest configuration uses both.
The gallery evidence exploited by the proposed terms is therefore independent of the session-level calibration rather than a byproduct of it.

\subsection{Gallery-size sensitivity}
\label{ssec:sweep}

\begin{table}[t]
\centering
\caption{Effect of the enrolled-gallery size $N$ on TST-Bench
(DIR@FAR=1\%,~[\%]). For $N{<}150$, mean$\pm$std over four random
speaker subsets; $N{=}150$ is the full gallery.}
\label{tab:sweep}
\setlength{\tabcolsep}{4pt}
\begin{tabular}{lcccc}
\toprule
$N$ & 10 & 30 & 75 & 150 \\
\midrule
Baseline     & 97.0$\pm$1.2 & 93.7$\pm$1.8 & 86.3$\pm$6.9 & 87.9 \\
$+$ Proposed & 96.2$\pm$1.4 & 95.6$\pm$1.2 & 90.5$\pm$5.2 & 91.1 \\
\bottomrule
\end{tabular}
\end{table}

Since the proposed terms derive all of their evidence from the gallery, their value should depend on how much structure the gallery offers.
Table~\ref{tab:sweep} varies the number of enrolled speakers.
Because accuracy at a given $N$ depends strongly on which speakers are drawn (baseline DIR of individual $N{=}75$ galleries ranges from $76.2\%$ to $91.5\%$), every $N{<}150$ condition is repeated over four random speaker subsets.
The average gain of the proposed calibration grows with the gallery.
At $N{=}10$ it is slightly negative, as roughly nine competitors provide neither a stable dispersion estimate nor a distinctive enrollment profile.
From $N{=}30$ onward the gain is positive for every sampled gallery, reaching about $2$\,pp at $N{=}30$ and $3$--$4$\,pp at $N{=}75$ and $N{=}150$.
The largest gains occur on the galleries where the baseline is weakest, and the calibration also reduces the spread across galleries.
The identical configuration is used at every $N$: the tail depth follows the square-root rule of Sec.~\ref{ssec:dispersion} and therefore adapts to the gallery size automatically.
The method is therefore best suited to the large-gallery regimes where false alarms are hardest to control~\cite{asimi2026}, which is precisely the regime that motivates TST deployments at scale.
Absolute numbers are not comparable across columns, as changing $N$ changes the target/non-target composition of the trials.

\section{Conclusion}
\label{sec:conclusion}

We analyzed how the enrollment gallery can be used to calibrate verification decisions in target speaker tagging.
Standard score normalization fails severely when the gallery serves as its cohort, and we traced the failure to per-speaker cohort statistics that reflect the clustering of enrollment conditions rather than impostor behavior.
The proposed gallery affinity verification exploits the same evidence through affine-invariant dispersion and profile-deviance terms.
It improves the tagging accuracy both with and without conventional score normalization, and it matches external-cohort normalization without any external resource, an attractive property for on-premises deployments where distributing large cohort sets is impractical.
A limitation is that the enrollment profiles assume a fixed gallery: enrolling or removing a speaker requires regenerating them, although only at enrollment time.
Future work includes relaxing this fixed-gallery assumption and extending the profile model to session-adaptive galleries.

\vfill\pagebreak

\bibliographystyle{IEEEbib}
\bibliography{mybib}

\end{document}